\documentclass{JHEP3} 

\usepackage{epsfig,multicol}
\usepackage{amsmath, amssymb}

\usepackage{concmath, palatino}

\newcommand\fverb{\setbox\pippobox=\hbox\bgroup\verb}
\newcommand\fverbdo{\egroup\medskip\noindent%
			\fbox{\unhbox\pippobox}\ }
\newcommand\fverbit{\egroup\item[\fbox{\unhbox\pippobox}]}
\newbox\pippobox

\newcommand{\be}{\begin{equation}} 
\newcommand{\ee}{\end{equation}}

\newcommand{\ba}{\begin{eqnarray}}
\newcommand{\ea}{\end{eqnarray}}

\title{On the supersymmetric vacua of the Veneziano-Wosiek model}

\author{Matteo Beccaria\\
  Dipartimento di Fisica, Universita' di Lecce,
  Via Arnesano, 73100 Lecce\\
  INFN, Sezione di Lecce\\
  E-mail: \email{matteo.beccaria@le.infn.it}}

\preprint{}

\abstract{
We study the supersymmetric vacua of the Veneziano-Wosiek model in sectors with fermion number $F=2, 4$
at finite 't Hooft coupling $\lambda$. We prove that for $F=2$ there are two zero energy vacua for $\lambda > \lambda_c = 1$
and none otherwise. We give the analytical expressions of both vacua. One of them was previously known, the second
one is obtained by solving the cohomology of the supersymmetric charges. 
At $F=4$ we compute the would-be supersymmetric vacua at high order in the the strong coupling expansion
and provide strong support to the conclusion that  $\lambda = 1$ is a critical point in this sector too.
It separates a strong coupling phase with two symmetric vacua from a weak coupling phase with positive spectrum.
}


\begin{document} 

\section{Introduction}
\label{Sec:Intro}

The study of quantum supersymmetric models for large number $N$ of degrees of freedom
has deep motivations in modern theoretical physics. Well known  examples are 
matrix model formulations of M-theory~\cite{Banks:1996vh} and AdS/CFT duality 
between ${\cal N}=4$ super Yang-Mills and type IIB superstring on $AdS_5\times S^5$~\cite{Minahan:2006sk}.

As illustrated in the recent review \cite{Wosiek:2006tc}, direct techniques are currently available
to analyze models in this class and a great deal of information can be obtained by combined 
analytical and numerical methods. These are based on an effective truncation of the state space
with the minor drawback of introducing a controlled, and eventually irrelevant, supersymmetry breaking.
Several examples at finite $N$ are discussed in~\cite{FiniteN}.

These methods can be extended to the most interesting limit $N\to\infty$ as explained in the beautiful series of 
papers~\cite{Veneziano:2005qs,Veneziano:2006dj,Veneziano:2006bx,Veneziano:2006cx}. Veneziano and Wosiek 
introduce a toy model of (non gauged) supersymmetric quantum mechanics at large $N$ and show that the $N\to\infty$ limit
can be described in terms of a planar Hamiltonian acting on single trace states. The dynamics in the planar limit is greatly simplified
and non-trivial analytical and numerical results can be obtained (see also~\cite{Onofri1,Onofri2} for related developments).

The Veneziano-Wosiek model is described in terms of $N\times N$ matrix fermion and boson 
creation/annihilation operators with (non trivial) algebra
\be
[a_{ij}, a^\dagger_{kl}] = \delta_{il}\delta_{jk}, \qquad
\{f_{ij}, f^\dagger_{kl}\} = \delta_{il}\delta_{jk}.
\ee
Supersymmetry is generated by the nilpotent charges
\be
Q = \mbox{Tr}[f\,a^\dagger\,(1+g\,a^\dagger)], \qquad
Q^\dagger = \mbox{Tr}[f^\dagger\,a\,(1+g\,a)],
\ee
where $g$ is a finite $N$ coupling constant. 
The supersymmetric Hamiltonian is 
\be
H = \{Q, Q^\dagger\}.
\ee
It commutes with $Q$, $Q^\dagger$, as well as with the additional operator $C = [Q^\dagger, Q]$ obeying $C^2=H^2$. 
The total fermion number $F = \mbox{Tr}(f^\dagger\,f)$ is conserved. The total boson number $B = \mbox{Tr}(a^\dagger\,a)$
varies by $\Delta B = 0, \pm 1$ under applications of $H$.

\bigskip
In the large $N$ limit, the Hamiltonian leaves invariant the subspace generated by single trace states of the form 
\be
|n_1, \dots, n_F\rangle = \mbox{Tr}[(a^\dagger)^{n_1}\,f^\dagger\cdots(a^\dagger)^{n_F}\,f^\dagger]\,|0\rangle.
\ee
In this limit, the natural coupling turns out to be the 't Hooft combination $\lambda = g^2\,N$ that will be kept fixed as $N\to\infty$.

\bigskip
The detailed analysis of the Veneziano-Wosiek model can be done in sectors with fixed $F$. At each $F$, one expects to find 
a certain number of supersymmetric vacua $b_F(\lambda)$. The positive energy states are paired with supersymmetric partners
in other sectors by the ladder action of the supersymmetry charges.
The analysis of~\cite{Veneziano:2005qs,Veneziano:2006dj,Veneziano:2006bx,Veneziano:2006cx}
investigates in details the $F=0,1,2,3$ cases at finite $\lambda$ and the general $F$ sector at infinite $\lambda$. 
The model turns out to be highly non trivial as we now summarize.

The simplest sectors $F=0,1$ can be treated analytically. The number of supersymmetric vacua is 
\be
b_0(\lambda) = \left\{\begin{array}{cc} 0, & \lambda < 1, \\ 1, & \lambda\ge 1\end{array}\right.,
\qquad
b_1(\lambda) = 0.
\ee
The positive energy levels are almost evenly spaced and perfectly paired between the $F=0,1$ sectors 
by the action of the supersymmetry charges. At the critical point $\lambda=1$ the spectrum collapses to zero. Across $\lambda=1$, 
there is a dynamical rearrangement of the SUSY multiplets. As a special feature, there is an exact strong/weak duality holding separately 
in both sectors.

The analysis of the next $F=2,3$ sectors is mainly numerical. At fixed $\lambda$, the total number
of boson excitations is truncated below a certain $B_{\rm max}$ and the Hamiltonian is diagonalized. The spectrum is then extrapolated
to $B_{\rm max}\to\infty$ limit. From the analysis, there are strong indications that 
\be
b_2(\lambda) = \left\{\begin{array}{cc} 0, & \lambda < \lambda_c, \\ 2, & \lambda\ge \lambda_c\end{array}\right.,
\qquad
b_3(\lambda) = 0,
\ee
with a critical $\lambda_c\simeq 1$. The positive energy levels are definitely not evenly spaced and they are only partially 
paired by the action of the supersymmetry charges. In other words, there are states with $F=3$ which are not annihilated by $Q^\dagger$. 
They should be paired with states with $F>3$ instead of $F=2$. There is no sign of any weak/strong coupling duality.
As a general fact, the convergence of the numerical extrapolation $B_{\rm max}\to\infty$ worsens as the critical point is approached.
In principle, this can be a practical difficulty in obtaining accurate estimates for $\lambda_c$.

The sectors with $F>3$ have been studied analytically by going to the extreme strong coupling limit $\lambda=\infty$
where the boson number $B$ is also conserved.
This limit is partially solvable by mapping the Veneziano-Wosiek model to other models, {\em i.e.} a gas of $q$-bosons and, notably, the integrable
XXZ spin $1/2$ chain with anisotropy $\Delta = -\frac{1}{2}$. Many exact properties of this spin chain are
known~\cite{CombinatoricsSpin}. In particular, it is possible to predict the number of supersymmetric vacua. It reads
\be
b_F(\infty) = \left\{\begin{array}{cc}
2, & \mbox{if}\ F\in 2 \mathbb{N},\ B = F\pm 1, \\
0, & \mbox{otherwise}.
\end{array}\right.
\ee
In the so-called  {\em magic case}, $ F\in 2 \mathbb{N}$ and $B = F\pm 1$, one can start from the two supersymmetric vacua at $\lambda=\infty$
and write formal power series in $1/\sqrt\lambda$ 
providing two zero energy states at finite $\lambda$. For one
of the two would-be vacua at $F=2$ it can be shown that the state is normalizable for $\lambda>1$. For the other finite $\lambda$ vacuum 
at $F=2$ and for the vacua at higher $F$ the coefficients of the strong coupling series are known in a quite implicit form. The 
extension of the strong coupling phase is thus unknown.

\bigskip
The supersymmetric vacua are states $\varphi$ obeying the two equations
\be
Q\,\varphi = Q^\dagger \,\varphi = 0.
\ee
Due to $Q^2 = (Q^\dagger)^2 = 0$, they compute the cohomology of $Q$ in the sector with $F$ fermions. 
One can expect to take some advantage in determining zero energy states by solving the above pair of equations instead of 
solving directly the equation $H\,\varphi=0$. 

In this paper, we follow this approach and extend the knowledge about  $b_F(\lambda)$ in two directions. First, at $F=2$, we determine the 
all-order expression of the second supersymmetric vacua proving rigorously that $\lambda_c = 1$ in that case too. This completes the analysis of the 
$F=2$ vacuum sector. Second, at $F=4$ we compute at high order the strong coupling expansion of the vacua providing strong and accurate numerical
support to the conclusion that again $\lambda_c = 1$. 

\bigskip
The plan of the paper is the following. In Sec.~(\ref{sec:planar}), we give all the relevant formulae to work out the
$N\to\infty$ planar limit of the Veneziano-Wosiek model. 
In Sec.~(\ref{sec:F2}), we provide the exact analytical expressions of the $F=2$ vacua.
In Sec.~(\ref{sec:F4}), we extend the analysis to the $F=4$ sector. Sec.~(\ref{sec:conclusions}) is devoted to conclusions.

\newpage

\section{The $N\to\infty$ limit of the Veneziano-Wosiek model}
\label{sec:planar}

\subsection{Hilbert space and norms}

In the $N\to\infty$ limit, the Hilbert space can be truncated to the $H$-invariant subspace generated by single trace 
states of the form 
\be
\label{eq:all}
|\mathbf{n}\rangle \equiv |n_1, \dots, n_F\rangle = \mbox{Tr}[(a^\dagger)^{n_1}\,f^\dagger\cdots(a^\dagger)^{n_F}\,f^\dagger]\,|0\rangle,\qquad n_i\in \mathbb{N}.
\ee
The basis states $|\mathbf{n}\rangle$ obey
\be
|S\,\mathbf{n}\rangle = (-1)^{F+1}\,|\mathbf{n}\rangle,\qquad 
\ee
where $S$ is the left shift operator acting on  $\mathbb{N}^F$ sequences as 
\be
S\,(n_1, \dots, n_F) = (n_2, \dots, n_F, n_1).
\ee
The Hilbert space in the sector with $F$ fermions is obtained by modding out the $\mathbb{Z}_F$ action of $S$. 
The $|\mathbf{n}\rangle$ states are not normalized. Their norm can be computed by applying the rules of planar calculus
as explained in~\cite{Veneziano:2005qs}. The result is 
\be
\Vert\ | \mathbf{n} \rangle \ \Vert^2 = d_{\mathbf{n}}\,N^{n_1+\cdots+n_F+F} + \mbox{subleading terms}.
\ee
The multiplicity $d_{\mathbf{n}}$ can be computed by the formula
\be
d_{\mathbf{n}} = \sum_{\ell=0}^{F-1}\,(-1)^{\ell\,(F+1)}\,\delta_{\mathbf{n}, S^\ell\,\mathbf{n}}.
\ee
Of course, null states with $d_\mathbf{n} = 0$ must be removed.
With self-explanatory notation, the first cases are
\ba
F=1 &:& d_a = 1, \nonumber \\
F=2 &:& d_{aa} = 0,\ \mbox{else}\ d_{ab} = 1, \nonumber \\
F=3 &:& d_{aaa} = 3,\ \mbox{else}\ d_{abc} = 1, \nonumber \\
F=4 &:& d_{aaaa} = 0,\ d_{abab}=2\  (a\neq b), \ \mbox{else}\ d_{abcd} = 1, \\
F=5 &:& d_{aaaaa} = 5,\ \mbox{else}\ d_{abcde} = 1,\nonumber \\
F=6 &:& d_{aaaaaa} = 0,\ d_{ababab}=3\ (a\neq b), \ d_{abcabc}=0, \ \mbox{else}\ d_{abcdef} = 1. \nonumber
\ea
In the following we shall arbitrarily choose a representative in each $S$-orbit and denote the 
resulting quotient Hilbert space as ${\cal H}_F$. A simple choice amounts to 
lexicographically order $\mathbf{n}$ and shift it by $S$ until the first element is $\min_i n_i$. In particular, $F=2$ basis states 
take the form $|n_1, n_2\rangle$ with $n_1<n_2$.

\subsection{Supersymmetry charges and Hamiltonian}

Let us split the supersymmetric charges in terms of operators with a definite variation of the boson number $B$. 
We write $Q = Q_1+Q_2$ with 
\ba
Q_1 &=& \phantom{g}\,\mbox{Tr}[f\,a^\dagger],\\
Q_2 &=& g\,\mbox{Tr}[f\,(a^\dagger)^2],
\ea
and similarly for the adjoint charges. We can go to the planar limit and remove all common powers of $N$. For instance, the norms
can be computed at leading order simply as
\be
\Vert\  | \mathbf{n} \rangle\ \Vert^2 = d_{\mathbf{n}}.
\ee
Then, by applying planar calculus, we obtain the following explicit formulas where the relevant coupling is indeed the 
't Hooft combination $\lambda$
\ba
Q_1^\dagger \,|n_1,\dots, n_F\rangle &=& \sum_{k=0}^{n_1-1} | k, n_1-k-1, n_2, \dots\rangle - \sum_{k=0}^{n_2-1} | n_1, k, n_2-k-1, n_3, \dots\rangle + \nonumber \\
 && \sum_{k=0}^{n_3-1} | n_1, n_2, k, n_3-k-1, n_4, \dots\rangle -\cdots.\\
\frac{1}{\sqrt\lambda}\,Q_2^\dagger \,|n_1,\dots, n_F\rangle &=& \sum_{k=0}^{n_1-2} | k, n_1-k-2, n_2, \dots\rangle - \sum_{k=0}^{n_2-2} | n_1, k, n_2-k-2, n_3, \dots\rangle + 
 \nonumber \\
 &&  \sum_{k=0}^{n_3-2} | n_1, n_2, k, n_3-k-2, n_4, \dots\rangle -\cdots ,
\ea
and 
\ba
Q_1\,|n_1,\dots, n_F\rangle &=& | n_1+n_2+1, n_3,  \dots\rangle - | n_1, n_2+n_3+1, n_4, \dots\rangle +  \\
 && | n_1, n_2, n_3+n_4+1, \dots\rangle -\cdots + (-1)^{F+1} | n_1 + n_F+1, n_2, \dots, n_{F-1}\rangle. \nonumber \\
\frac{1}{\sqrt\lambda}\,Q_2\,|n_1,\dots, n_F\rangle &=& | n_1+n_2+2, n_3,  \dots\rangle - | n_1, n_2+n_3+2, n_4, \dots\rangle +  \\
 &&  | n_1, n_2, n_3+n_4+2, \dots\rangle -\cdots + (-1)^{F+1} | n_1 + n_F+2, n_2, \dots, n_{F-1}\rangle . \nonumber
\ea
In principle, one could also write an explicit expression for the Hamiltonian. We shall not need it, apart from the $F=2$ case. The expression
is a bit involved and reads
\ba
H\,|n_1, n_2\rangle &=& [(n_1+n_2+2)(1+\lambda)-\lambda(2-\delta_{n_1, 0}+2\,\delta_{n_1+1, n_2})]\,|n_1, n_2\rangle + \nonumber \\
&& + \sqrt\lambda\,[(n_1+2)\,|n_1+1, n_2\rangle + (n_1+1)\,|n_1-1, n_2\rangle + \\
&& \ \ \ \ \ \ \ \  + (n_2+2)\,|n_1, n_2+1\rangle + (n_2+1)\,|n_1, n_2-1\rangle] + \nonumber\\
&& + 2\,\lambda\,[(1-\delta_{n_1+1, n_2})\,|n_1+1, n_2-1\rangle+(1-\delta_{n_1, n_2+1})\,|n_1-1, n_2+1\rangle]. \nonumber
\ea
Notice that it involves $\lambda^0, \lambda^{1/2}, \lambda^1$ contributions.

\newpage
\section{Analytical supersymmetric vacua at $F=2$}
\label{sec:F2}

From the explicit form of $Q$ we obtain $Q\,{\cal H}_2 = 0$. This is in agreement with the SUSY multiplet structure of the $F=0,1$ sectors.
As we recalled in the Introduction, all states in ${\cal H}_1$ are in the image of $Q^\dagger$. Hence, given $\varphi\in {\cal H}_2$
we have $Q\,\varphi = Q^\dagger \,\eta$ for some $\eta\in {\cal H}_0$. Hence, 
\be
Q^\dagger\,Q\,\varphi = (Q^\dagger)^2\,\eta = 0.
\ee
This means that $\Vert\,Q\,\varphi\,\Vert^2 = 0$, or $Q\,\varphi = 0$. Therefore, the zero energy states in ${\cal H}_2$ are the solutions
to the single equation
\be
\label{eq:eq1}
Q^\dagger \,\varphi = 0.
\ee
To solve it, it is convenient to filter $\varphi$ according to the boson number $B$. We write
\be
\varphi = \sum_{B\ge 1}\varphi_B,
\ee
where $\varphi_B$ has $B$ boson excitations (there are no states with $B=1$ in ${\cal H}_2$).
Replacing in Eq.~(\ref{eq:eq1}), we find
\be
\label{eq:eq2}
Q_1^\dagger \,\varphi_B + Q_2^\dagger \, \varphi_{B+1} = 0.
\ee
This equation determines $\varphi_{B+1}$ in terms of $\varphi_B$ as we now show. 

\bigskip\bigskip\noindent
At $B=1$, Eq.~(\ref{eq:eq2}) has the unique solution
\be
\varphi_1 = \alpha\,|0,1\rangle.
\ee
At $B=2$ we also obtain a unique solution
\be
\varphi_2 = -\frac{\alpha}{\sqrt\lambda}\,|0,2\rangle.
\ee
The most general solution at $B=3$ is 
\be
\varphi_3 = \frac{\alpha}{\lambda}\,|0,3\rangle + \beta\left(-\frac{1}{2}\,|0,3\rangle + |1, 2\rangle\right).
\ee
The existence of an additional arbitrary constant is consistent with the $\lambda=\infty$ analysis which also 
predicts that for $B>3$ there should be no additional arbitrary constants. Taking $\beta=0$ and $\alpha=1$ and iterating we 
easily obtain the explicit expression of the first supersymmetric vacuum
\be
\varphi^{(1)} = \sum_{B\ge 1} \frac{(-1)^{B+1}}{(\sqrt\lambda)^{B-1}}\,|0, B\rangle.
\ee
This is normalizable for $\lambda>1$ and is the vacuum already found in~\cite{Veneziano:2006bx}. 

\noindent
Taking $\alpha=0$ and $\beta=1$
we obtain the following expressions for the various terms of the second vacuum
\ba
\varphi_1 &=& \varphi_2 = 0, \\
\varphi_3 &=& -\frac{1}{2}\,|0,3\rangle + |1, 2\rangle, \\
\varphi_4 &=& \frac{1}{\lambda^{1/2}}\left(|0,4\rangle-\frac{5}{4}\,|1,3\rangle\right),\\
\varphi_5 &=& \frac{1}{\lambda}\left(-\frac{17}{12}\, |0,4\rangle+\frac{5}{4}\,|1,4\rangle+\frac{5}{12}\,|2,3\rangle\right),\\
\varphi_6 &=& \frac{1}{\lambda^{3/2}}\left(\frac{7}{4}\, |0,6\rangle-\frac{7}{6}\,|1,5\rangle-\frac{7}{12}\,|2,4\rangle\right),
\ea
and so on. Working out several additional levels, we are led to the following conjectured expressions
\be
\varphi_B = \frac{1}{(\sqrt\lambda)^{B-3}}\sum_{n=0}^{N_B} c_{B, n}\ |n, B-n\rangle,\qquad N_B = \left\{
\begin{array}{cc} 
\displaystyle \frac{B-1}{2}, & B\in 2\,\mathbb{N}+1, \\ \\
\displaystyle \frac{B}{2}-1, & B\in 2\,\mathbb{N}, \\
\end{array} \right. ,
\ee
with the coefficients
\ba
\label{eq:conj}
c_{B, 0} &=& (-1)^B\,\frac{7}{2}\,\frac{(B-1)(B-2)(B+12)}{(B+2)(B+3)(B+4)},\qquad B\ge 1, \\
c_{B, n} &=& -210\,(-1)^B\frac{B-2\,n}{(B+2)(B+3)(B+4)},\qquad B\ge 2\,n. \nonumber
\ea

\bigskip
We now prove that Eqs.~(\ref{eq:conj}) are indeed the unique solution of the basic equation Eq.~(\ref{eq:eq1}).
This is not as trivial as it could appear at first sight. Indeed, acting with $Q^\dagger$ produces states in ${\cal H}_3$ which are not
necessarily in canonical order and the check is annoying. An alternative proof exploits the more complicated equation $H\,\varphi=0$
which is more convenient in order to check the conjecture. Applying $H$ to $\varphi$ we obtain
\ba
\lefteqn{
\sum_{B, n}\frac{c_{B, n}}{(\sqrt\lambda)^B}\left\{\ 
[(B+2)(1+\lambda)-\lambda\,(2-\delta_{n,0}-2\,\delta_{n+1, B-n})]\,|n, B-n\rangle + \right. } && \nonumber \\
&& \sqrt\lambda[ (n+2) |n+1, B-n\rangle + (B-n+2) |n, B-n+1\rangle + \\
&& + (n+1) |n-1, B-n\rangle + (B-n+1) | n, B-n-1\rangle] + \nonumber \\
&& \left. + 2\,\lambda[(1-\delta_{n+1, B-n}) | n+1, B-n-1\rangle + (1-\delta_{n, B-n+1}) | n-1, B-n+1\rangle]\  \right\}. \nonumber
\ea
This gives terms proportional to  $\lambda^0$ or $\lambda$ leading to the two recursion equations
\be
(B+\delta_{n, 0}-2\,\delta_{n+1, B-n})\, c_{B, n} + (n+1)\,c_{B-1, n-1} + (B-n+1)\, c_{B-1, n} +
\ee
$$
+  2\,(1-\delta_{n, B-n+1})\, c_{B, n-1}
 + 2\,(1-\delta_{n+1, B-n})\, c_{B, n+1} = 0,
$$
\be
(B+2)\, c_{B, n} + (n+2)\, c_{B+1, n+1} + (B-n+2)\, c_{B+1, n} = 0.
\ee
In the above two equations one has to set $c_{B, n}=0$ if $n$ does not obey $0\le n\le N_B$. The boundary conditions
that fix uniquely the solution to the recursion are
\ba
c_{B, n} &=& 0, \qquad B = 1, 2, \\
c_{3, 0} &=& -\frac{1}{2}, \\
c_{3, 1} &=& 1.
\ea
It is an easy check to verify that Eq.~(\ref{eq:conj}) indeed solve the above recursion with the assigned boundary conditions.

\bigskip
Thus, we have found the explicit expression for the second vacuum and it reads
\be
\varphi^{(2)} =\sum_{B\ge 3} \frac{1}{\lambda^{(B-3)/2}}\frac{(-1)^B}{(B+2)(B+3)(B+4)} \left\{
\frac{7}{2}(B-1)(B-2)(B+12) | 0, B\rangle\right. 
\ee
$$
\left.  -210\sum_{n=1}^{N_B}(B-2\,n)\,|n, B-n\rangle \right\}.
$$
Evaluating the norm using the relation (holding $\forall B\in\mathbb{N}$)
\be
\sum_{n=1}^{N_B}(B-2\,n)^2 = \frac{1}{6}\,B\,(B-1)\,(B-2),
\ee
we find
\be
\Vert\ \varphi^{(2)}\ \Vert^2 = \frac{49}{4}\,\sum_{B\ge 3} \left(\frac{1}{\lambda}\right)^{B-3}\ \frac{(B-1)\,(B-2)\,(B^3+19\,B^2+36\,B+144)}{(B+2)(B+3)^2(B+4)^2}
\ee
The series converges for $x = \frac{1}{\lambda} < 1$ and can also be resummed with the explicit result
\ba
\Vert\ \varphi^{(2)}\ \Vert^2 &=& -\frac{49}{4}\frac{x^4+20\,x^3-1610\, x^2-8670\, x+10260}{x^6\,(x-1)} + \\
&& -735\frac{(x-1)(111+7\,x)}{x^7}\,\log(1-x) - 44100\frac{x+1}{x^7}\,\mbox{Li}_2(x),
\ea
where $\mbox{Li}_2(x)$ is the dilogarithm function. From this expression, we extract the singular behavior in the $x\to 1^-$ limit
(the prefactor is an arbitrary normalization constant following our choice $\beta=1$)
\be
\Vert\ \varphi^{(2)}\ \Vert^2 = \frac{49}{4}\,\frac{1}{1-x} + \dots.
\ee
In conclusion, we have shown that the cohomology of $Q$ at $F=2$ has two solutions both normalizable for $\lambda>\lambda_c=1$ only, {\em i.e.}
\be
b_2(\lambda) = \left\{\begin{array}{cc} 0, & \lambda < 1, \\ 2, & \lambda\ge 1\end{array}\right. .
\ee

\newpage
\section{Results at $F=4$}
\label{sec:F4}

\subsection{A view to the spectrum}

The sector $F=4$ is much more complicated. As a first step, we have diagonalized $H$ up to $B_{\rm max}=24$ to have a feeling about would-be zero modes
in the $B_{\rm max}\to\infty$ limit. The smallest 6 levels are shown in 
Fig.~(\ref{fig:spectrum}) for the values $\lambda = 0.5$, $1$, $1.2$, and $4$. For $\lambda = 0.5$ and $\lambda = 4$ it seems quite clear
that there are respectively 0 and 2 supersymmetric vacua. At $\lambda=1$ it is plausible that all levels are converging to zero in agreement with the
reasonable conjecture that the critical point is again $\lambda_c = 1$. However, the estimate of $\lambda_c$ is difficult at these values of $B_{\rm max}$
as illustrated by the inset at $\lambda=1.2$. Here a clean stabilization as for $\lambda=4$ would require quite  a larger $B_{\rm max}$.
If we do not want to push further the numerical diagonalization, it seems mandatory
to find an alternative determination of the critical point. 

Are the methods exploited at $F=2$ applicable ? In the next Section we shall address this question discussing
some difficulties and their (numerical) resolution.

\subsection{Strong coupling expansion of $Q$ cohomology}

At $F=4$ we find zero energy states by imposing the full set of cohomological equations 
\be
\label{eq:eq3}
Q\,\varphi = Q^\dagger\,\varphi = 0.
\ee
The general solution is quite complicated compared to the case $F=2$ and the solution does not organize well in powers of $\lambda$.
The reason is that $Q^\dagger\,\varphi$ produces a series in descending powers of $\sqrt\lambda$ as at $F=2$. However, the 
equation $Q\,\varphi = 0$ has the opposite behavior.

We can bypass this problem recalling that, after all, we are interested in the determination of the 
convergence radius of the strong coupling expansion. Thus, we try to solve Eqs.~(\ref{eq:eq3}) by making from start 
the Ansatz
\be
\varphi = \sum_{n=0}^\infty \lambda^{-n/2}\,\sum_{B=1}^\infty \varphi_{n, B},
\ee
where, as indicated, $\varphi_{n, B}$ is a state with boson number $B$. Removing the $\sqrt\lambda$ factors in $Q_2$ and $Q_2^\dagger$ 
we have to solve the equations
\ba
\label{eq:eq4}
Q_2\,\varphi_{n, B} + Q_1\,\varphi_{n-1, B+1} &=& 0, \\
Q_2^\dagger\,\varphi_{n, B} + Q_1^\dagger\,\varphi_{n-1, B-1} &=& 0. \nonumber
\ea
It is easy to check that these equations are compatible and admit a unique solution for the $n$-th order in terms of $\varphi$ computed
at $(n-1)$-th order. This is true with the exception of those values of $B$ where the operator $Q_2$ has non empty cohomology. However, the cohomology
of $Q_2$ is given by the zero energy states of the $\lambda=\infty$ Veneziano-Wosiek model which is known. It contains a state 
at each $B = F\pm 1$, here $B=3, 5$. The explicit zero modes (with an arbitrary normalization) are 
\ba
|\eta_3\rangle &=& |0, 0, 0, 3\rangle - 3\, |0, 0, 1, 2\rangle + 3\, |0, 0, 2, 1\rangle + 4\, |0, 1, 0, 2\rangle - 7\, |0, 1, 1, 1\rangle, \\
|\eta_5\rangle &=& -|0, 0, 0, 5\rangle + 4\, |0, 0, 1, 4\rangle - 7\, |0, 0, 2, 3\rangle + 
  7 \,|0, 0, 3, 2\rangle - 4 \,|0, 0, 4, 1\rangle  + \\
&& - 6 \,|0, 1, 0, 4\rangle + 
  17 \,|0, 1, 1, 3\rangle - 21 \,|0, 1, 2, 2\rangle + 14 \,|0, 1, 3, 1\rangle - 
  12 \,|0, 2, 0, 3\rangle + \nonumber \\
&& + 25 \,|0, 2, 1, 2\rangle - 21 \,|0, 2, 2, 1\rangle + 
  17 \,|0, 3, 1, 1\rangle - 42 \,|1, 1, 1, 2\rangle. \nonumber 
\ea
Let us discuss in some details the solution which reduces at $\lambda=\infty$ to $|\eta_3\rangle$. The other case is completely similar.
We start with 
\be
\varphi_0 = \sum_B\,\varphi_{0, B} = |\eta_3\rangle.
\ee
Then, we solve at each $B$ Eq.~(\ref{eq:eq4}) with $n=1$. Of course, there is a maximum $B$ beyond which we do not have non vanishing 
solutions for $\varphi_{1, B}$. The procedure is iterated. The solution of Eq.~(\ref{eq:eq4}) is always unique with the exception of the 
cases $B=3, 5$ where we can add to $\varphi_{n, B}$ an arbitrary constant $\alpha$ times $|\eta_B\rangle$. The general solution can always be put in the form 
\ba
\varphi_{n, 3} = \varphi_{n, 3}^{\rm inhom} + \alpha\, |0, 1, 1, 1\rangle, \\
\varphi_{n, 5} = \varphi_{n, 5}^{\rm inhom} + \alpha\, |1, 1, 1, 2\rangle.
\ea
In other words, the inhomogeneous piece of the solution does not have contributions from the states $|0, 1, 1, 1\rangle$ and $|1, 1, 1, 2\rangle$
which come totally from the zero modes. We arbitrarily set $\alpha=0$ to fix the zero mode contributions. Other choices are possible, but do not
change the convergence properties of the strong coupling expansion.

\bigskip
\noindent
The explicit expression of $\varphi_{n, B}$ is quite complicated and unfortunately we did not succeed in finding a closed formula. 
However one can try to estimate the convergence radius from a study of the strong coupling series. To this aim, we have evaluated
the norm of the would-be vacuum by working out the terms of 
\be
\Vert\ \varphi\ \Vert^2 = \sum_{n=0}^\infty \frac{a_n}{\lambda^n},
\ee
up to $n=24$.  After normalization, the first terms read
\ba
\Vert\ \varphi\ \Vert^2 &=&
1 + 
\frac{33163}{13272}\,\frac{1}{\lambda} + 
\frac{3606544643}{777060669}\,\frac{1}{\lambda^2}+
\frac{6669989903943227}{891227976610818}\,\frac{1}{\lambda^3}+ \nonumber \\
&& + \frac{2155907292859955213802297145858}{195526455552229171879002565071}\,\frac{1}{\lambda^4}+ \nonumber \\
&& + 
\frac{5974912975520703560602997490582425731877554327739}{393252674274544022631209524089738568938469988544}\,\frac{1}{\lambda^5} + \cdots. \nonumber
\ea
The convergence radius can be estimated by the ratio test or, better, by means of improved recurrent estimators
in the spirit of~\cite{radius}. In other words, we compute the sequences
\ba
R^{(0)}_n &=& \frac{a_{n+1}}{a_n}, \\
R^{(1)}_n &=& n\,\frac{a_{n+1}}{a_n}-(n-1)\,\frac{a_n}{a_{n-1}}, \\
R^{(2)}_n &=& \frac{1}{2}\left[n^2\,\frac{a_{n+1}}{a_n}-2\,(n-1)^2\,\frac{a_n}{a_{n-1}}+ (n-2)^2\,\frac{a_{n-1}}{a_{n-2}}\right], \\
R^{(3)}_n &=& \frac{1}{6}\left[n^3\,\frac{a_{n+1}}{a_n}-3\,(n-1)^3\,\frac{a_n}{a_{n-1}}+ 3\,(n-2)^3\,\frac{a_{n-1}}{a_{n-2}}
-(n-3)^3\,\frac{a_{n-2}}{a_{n-3}}\right], \\
\overline{R}^{(3)}_n &=& \frac{1}{2}(R^{(3)}_n+R^{(3)}_{n+1}).
\ea
Fig.~(\ref{fig:ratio}) show the results obtained with $R^{(0)}$, $R^{(1)}$, $R^{(2)}$, and $\overline R^{(3)}$.
As one can see, the ratio test (sequence $R^{(0)}$) is poorly useful in determining $\lambda_c$. Instead,
the higher order estimators converge more and more quickly to a $\lambda_c$ that can be estimated to be 
\be
\lambda_c = 1.000(1).
\ee
This result shows that there is a supersymmetric vacuum extending up to $\lambda_c\simeq 1$ at $F=4$. The same procedure
can be started from the strong coupling vacuum at $B=5$, repeating the construction and removing the component along the
first vacuum in order to enforce orthogonality. The numerics is less clean, but fully consistent with the above estimate.

\noindent
Thus, we have provided strong support to the conclusion that
\be
b_4(\lambda) = \left\{\begin{array}{cc} 0, & \lambda < 1, \\ 2, & \lambda\ge 1\end{array}\right. .
\ee
It is clear that the methods described in this section can be extended to larger $F$ with no additional difficulties.

\section{Conclusions}
\label{sec:conclusions}

The Veneziano-Wosiek model is a surprisingly rich toy model where quantum supersymmetry at large $N$ can be investigated.
As is usual in supersymmetry, a lot of information is already contained in the most basic question, the dimension of 
the vacuum sector, the integer number $b_F(\lambda)$. In this paper, we have extended the known results for $F=0, 1, 2, 3$
providing new analytical results at $F=2$ and $F=4$. Our results support the conjecture that the two 
strong coupling supersymmetric vacua existing for even $F\ge 2$ 
can be analytically continued up to the critical value $\lambda_c=1$ in all fermion sectors.

Most interestingly, the Veneziano-Wosiek model is known to have some intriguing connection with combinatorial problems
as discussed in~\cite{Onofri1}. This fact is well established in the extreme strong coupling limit. The mapping to the 
XXZ spin chain permits to extend to the Veneziano-Wosiek model several number-theoretical facts~\cite{CombinatoricsSpin} 
recently exploited in the context of Alternating sign matrix conjectures~\cite{CombinatoricsASM}.
Similar relations between supersymmetric models and combinatorics are actually not new as discussed in the SUSY algebra non-linear
realizations discussed in~\cite{CombinatoricsFermion} and also related to the XXZ chain at the peculiar anisotropy $\Delta = -\frac{1}{2}$.

What is somewhat surprising is the fact that hidden combinatorial facts could be at work even at finite coupling.
The search for supersymmetric vacua in the $F=2$ sector described in this paper has been achieved due to the ability of
guessing the solution of a complicated recursion problem. As soon as a guess is proposed, it can be checked with minor effort.
However the guess itself was not trivial. Actually, we could find it by searching within suitable 
classes of rational sequences arising precisely in typical combinatorial problems~\cite{determinant}.

\acknowledgments
We thank J. Wosiek for communications about his numerical results on the model and G. F. De Angelis for conversations.

\newpage
\FIGURE{
\epsfig{file=spectrum.F4.eps,width=14cm}\bigskip\bigskip
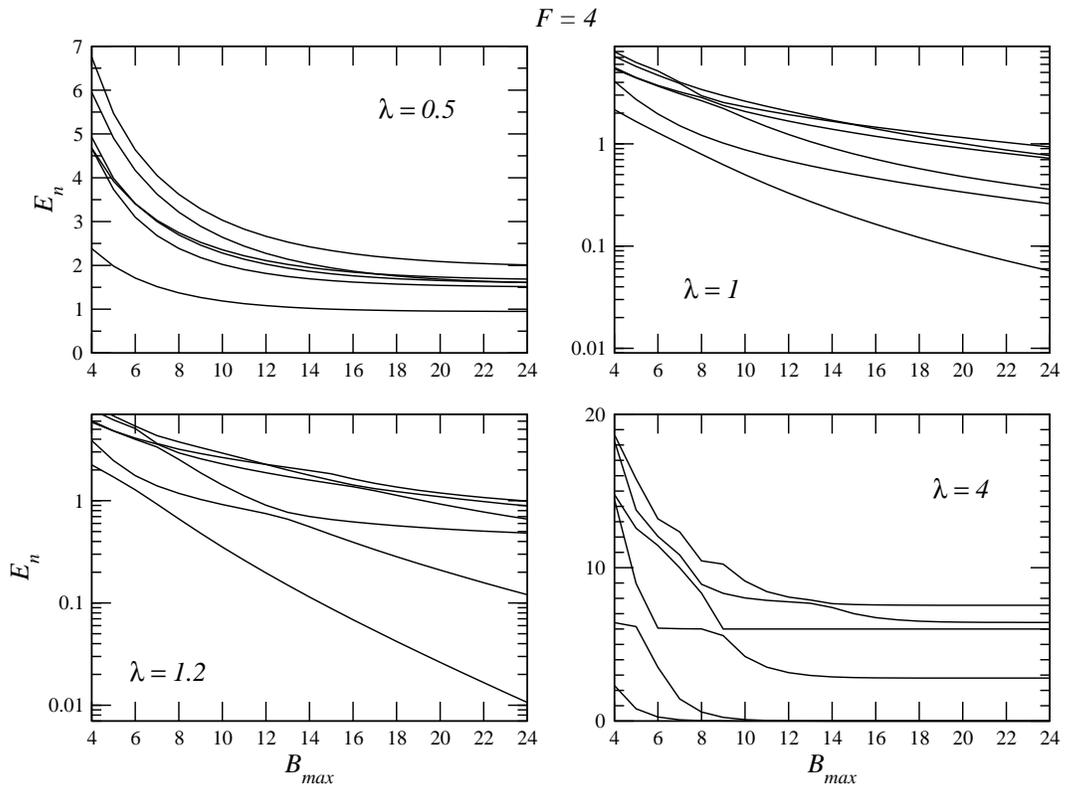
\caption{Smallest 6 energy levels at $F=4$ and various couplings $\lambda$ as functions of the upper limit on the boson 
number $B\le B_{\rm max}$.}
\label{fig:spectrum}
}

\newpage
\FIGURE{
\epsfig{file=ratio.FZ3.eps,width=14cm}\bigskip\bigskip
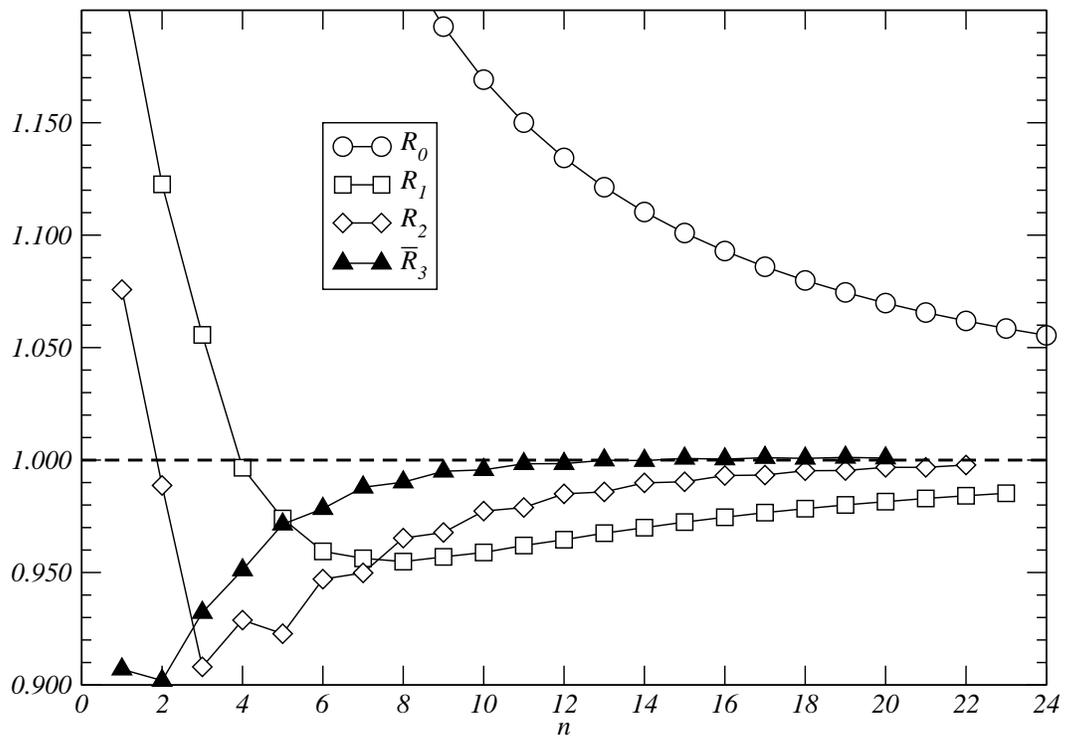
\caption{Estimate of $\lambda_c$ from ratio and recurrent ratio-like tests in the $F=4$ case.}
\label{fig:ratio}
}

\end{document}